\documentstyle[prl,twocolumn,epsf,floats,aps]{revtex}

\begin{document}
\draft

\twocolumn[\hsize\textwidth\columnwidth\hsize\csname
@twocolumnfalse\endcsname

\title{Comment on ``Cooper instability of composite fermions''}
\author{N. Read}
\address{Department of Physics, Yale University, P.O. Box 208120,
New Haven, CT 06520-8120}
\date{October 4, 2000}
\maketitle

\begin{abstract}
We comment on the recent paper by Scarola, Park, and Jain
{[}Nature {\bf 406}, 863 (2000){]} on a trial wavefunction
calculation of pairing in a fractional quantum Hall system at
$\nu=5/2$. We point out two errors that invalidate the claimed
calculations of a binding energy for Cooper pairs and of an energy
gap for charged excitations.
\end{abstract}

\vspace{0.1in}
]

The old problem of the incompressible fractional quantum Hall
state observed at filling factor $\nu=5/2$ \cite{willett} has
received renewed attention recently. In particular, a paper by
Scarola, Park, and Jain (SPJ) \cite{spj} presents a numerical
calculation based on trial wavefunctions which finds that
composite fermions \cite{jain,hlr} bind to form Cooper pairs at
half-filling of the first excited (or ${\cal N}=1$) Landau level
($\nu=5/2$), but not at half-filling of the lowest (${\cal N}=0$)
Landau level ($\nu=1/2$).

First, earlier work on this problem along similar lines should be
properly pointed out. The idea that composite (or neutral)
fermions at even denominator filling factors (such as $1/2$,
$5/2$) can form an incompressible state by pairing as in
Bardeen-Cooper-Schrieffer (BCS) theory \cite{bcs} was pointed out
in Ref.\ \cite{mr}. The existing spin-singlet Haldane-Rezayi state
\cite{hr} was interpreted in this way, and a spin-polarized state,
termed the Pfaffian state, was constructed as another example
\cite{mr}. Later, Greiter {\it et al.} explored this idea further,
and suggested that the Pfaffian state might apply to $\nu=5/2$
\cite{gww}. The incompressibility of the paired state may be
rephrased by invoking the ``Meissner effect'', see for example
Ref.\ \cite{read94}. This is of course a ``composite boson'' type
of explanation for incompressibility \cite{g,gm,read87,zhk}.

In the early days, mainly because of Ref.\ \cite{eisenstein}, it
was assumed in most theoretical work (an exception being the
suggestion in Ref.\ \cite{gww}) that the ${\cal N}=1$ Landau level
is spin-unpolarized at $\nu=5/2$---including the earlier work,
Ref.\ \cite{belkhir}. Recently, convincing evidence that the
${\cal N}=1$ Landau level is polarized in these states, even when
the Zeeman energy is turned off, has been presented in
diagonalization studies \cite{morf}. This is assumed without
comment in SPJ. The recent, very extensive, numerical studies find
that the ground state at $5/2$, unlike that at $1/2$, is an
incompressible state with the quantum numbers of, and a large
overlap with, the Pfaffian state (or its projection to a
particle-hole symmetric state), and that phase transitions to
compressible states occur when the interaction is modified away
from the Coulomb interaction \cite{morf,rh}. Thus, the evidence
{\em already} favored a paired state of composite fermions at
$\nu=5/2$, whereas for $\nu=1/2$, similar studies showed
Fermi-liquid-like behavior \cite{rr,rh}, consistent with Ref.\
\cite{hlr}.

Further, for $\nu=5/2$, Morf \cite{morf} obtained a gap for
charged excitations (well-separated quasiparticle and quasihole)
of $\Delta=0.025$ in units of $e^2/\l_0$; SPJ find a number 5
times smaller, but do not make the comparison with Morf's result.
Experimental results \cite{jim,pan} are smaller still, about a
factor of 5 smaller than the SPJ result in the case of Ref.\
\cite{pan}.

Technically, SPJ use trial wavefunctions for $N$ electrons on the
sphere at magnetic flux $N_\phi=2(N-1)$, consisting of Slater
determinants of spherical harmonics (representing fermions in zero
net magnetic field) times a Laughlin-Jastrow factor, projected to
the lowest Landau level, to represent ground states at $\nu=1/2$
(the $\nu=5/2$ problem can be mapped to this, provided the Coulomb
interaction appropriate to the ${\cal N}=1$  Landau level is used
as the Hamiltonian). The first use of these functions for the
$\nu=1/2$ problem at this sequence of system sizes was in Ref.\
\cite{rr}. In a comment, Jain claimed that the $\nu=1/2$ state
(with ${\cal N}=0$ Coulomb interaction) in the thermodynamic limit
could not be approached in this way \cite{jainprlcom}. Also, the
consistency of the angular momentum of the ground states for each
$N$ with Hund's rule was a result in Ref.\ \cite{rr}; this was
dismissed as ``wild fluctuations'' \cite{jainprlcom}, yet all
these results are now accepted in SPJ, without citing the earlier
discussion. (The corresponding energies were used to estimate the
composite fermion effective mass at $1/2$ by Morf and
d'Ambrumenil, who also examined $\nu=5/2$ \cite{md}.) We do not
believe that, for the Fermi-liquid state, the thermodynamic limit
must be approached along the subsequence $N=n^2$ (where filled
shells occur \cite{rr}) as SPJ suggest. The reason is that the
fluctuation in the total angular momentum of the ground state
caused by partially-filled shells of composite fermions is
relatively small (only order $\sqrt{N}$ particles are involved),
and correlation functions of local operators at fixed separations
will be insensitive to this fluctuation as $N\to\infty$.

Now we turn to more substantive errors in SPJ. The first concerns
the order of magnitude of an interaction matrix element. The trial
wavefunctions are supposed to represent two fermions in the
$n+1$th shell, each of angular momentum $l=n$, and the lower
filled shells form a Fermi sea \cite{rr}. The distinct states can
be labeled by their total angular momentum $L$, which is $1$, $3$,
\ldots, $2n-1$. SPJ interpret the difference in energy $E_L$
(calculated by Monte Carlo) between the $L=2n-1$ and $L=1$ trial
states as an interaction energy for the two added fermions. When
$E_1-E_{2n-1}$ is negative, the pair are trying to form a Cooper
pair. We will consider this model problem of two fermions outside
a Fermi sea (in zero magnetic field) on a sphere, where they
interact only with each other, not with the Fermi sea (similar to
the ``Cooper problem'' \cite{cooper}). In the model, all the
states of the two fermions each with angular momentum $n$ are
degenerate, in the absence of the interaction, and are split from
other $l$ values by an effective kinetic energy.

The problem arises because each $L$ eigenstate of two fermions,
each with angular momentum $n$, is a linear combination of only of
order $\sqrt{N}$ states (since $N=n^2+2$). Because the spherical
harmonics of angular momentum $n$ are extended over the sphere,
this is too few to obtain a bound state wavefunction of fixed size
as $N\to\infty$. The smallest wavepacket that can be constructed
covers an area of order $\sqrt{N}$ in the {\em relative}
coordinate (the unit length is the magnetic length of the original
quantum Hall problem, and the system area $4\pi R^2$ is of order
$N$). A true bound state would have a size of order one; for this,
order $N$ basis states are needed.

Related to this, the expectation value of the interaction energy
of two particles in such a state built from only order $\sqrt{N}$
basis states is of order $1/\sqrt{N}$. For any physically-sensible
interaction, including long-range interactions such as the Coulomb
interaction, the matrix elements of the interaction between
normalized, extended, basis states of fixed single-particle {\em
linear} momentum, are of order $1/N$ (or $1/R^2$). Here we take
the $N\to\infty$ limit at fixed linear momenta of the in- and
out-going particles; for the sphere, linear momentum $k$ (as in
the plane) corresponds to $l/R$. For $l$ of order $\sqrt N\propto
R$ (to maintain fixed density of particles), this is of order one,
as we expect since it is of order the Fermi wavevector. This
behavior of the matrix elements is familiar for plane waves and
periodic boundary conditions in flat space (see e.g. Ref.\
\cite{fetter} for the three-dimensional analog), but is also true
for other cases including the sphere. When these matrix elements
are used to calculate the expectation of the interaction for two
particles each of angular momentum $l=n$ in the state of total
angular momentum $L=1$ (or any other value up to the limit
$2n-1$), the result is of order $1/\sqrt{N}$, because there are
order $N$ diagonal and off-diagonal terms, and we divide by
$\sqrt{N}$ to normalize the state. (The Clebsch-Gordan
coefficients used in forming the $L=1$ state scale uniformly as
$N^{-1/4}$, as one can see from explicit expressions \cite{ll}.)
The same dependence can be obtained easily on the plane by
restricting each particle to a set of $\sqrt{N}$ single-particle
states $k$ just outside the Fermi wave vector, and pairing $k$
with $-k$.

Order $N$ basis states would be needed to obtain an expectation
value of the interaction of order 1. Since, in the model, such
states are not degenerate in the absence of interaction, and the
ground state is not determined solely by symmetry as it was
before, a correct calculation then involves solving the
Schr\"{o}dinger equation, as first done by Cooper \cite{cooper}
(and easily adapted to the present case). If it is claimed that,
contrary to our analysis, the expectation of the interaction in
the state in the smaller basis set is independent of system size,
then this gives a contradiction with Cooper's result for weak
coupling, which was that the binding energy is exponentially small
for weak coupling, not linear in the coupling as it is in the
smaller basis set.

SPJ take their numerical data and plot it versus $1/N$, using a
linear extrapolation to a nonzero value. We find that, within the
interpretation used, this is inappropriate, since as an
interaction matrix element it should approach zero as
$1/\sqrt{N}$. This is the first error. In the plots of SPJ, the
data does appear linear in $1/N$. This must mean either that it
will eventually turn over to $1/\sqrt{N}$, or that the trial
states should not be interpreted as two fermions just outside the
Fermi sea as in Refs.\ \cite{rr,spj} and the analysis here.

The second error comes when SPJ compare their ``binding energy''
with the gap for charged excitations in the quantized Hall system
at $\nu=5/2$. We leave aside the obvious question of whether such
a binding energy would give accurately the energy gap in the
many-particle ground state. More importantly, the fermions
obtained by breaking a pair are neutral \cite{mr}; this is true
even in the Fermi liquid state without pairing
\cite{read94,sm,dhlee,ph,read98}. The charged excitations are
vortices which, because of pairing, effectively contain a half
quantum of flux each, and so have charge $\pm1/4$ (in electronic
units) at $\nu=5/2$ \cite{mr}. There appears to be no expected
relation between the excitation energy gaps for these two very
different types of excitations. This again invalidates the
comparison with experimental results.

Another way to understand this point is in terms of what change
must be made in the ground-state quantum numbers to create certain
excitations. SPJ add two particles to a filled shell, staying on
the sequence $N_\phi=2(N-1)$, and this means their excitation,
when a pair is broken, is neutral. Of course, such a state may
contain charged particles of opposite charges, but this is not
obvious. For this reason, charged excitations are usually obtained
by adding or subtracting flux or particles so as to leave a given
sequence. This is a well-known technique, used for example by Morf
to obtain a charge-excitation gap for the $5/2$ state \cite{morf}.
Since the elementary charged excitations contain a half flux
quantum in the present case, Morf divides the number obtained that
way by 2, after subtracting the interaction energy for two $\pm
1/4$ charges and extrapolating to the thermodynamic limit, to
obtain his result $0.025$.

In spite of these errors, the calculation by SPJ is still an
interesting trial wavefunction calculation of something like an
interaction matrix element, which is attractive in the ${\cal
N}=1$, but not in the ${\cal N}=0$, Landau level, a fact
consistent with earlier results about the different spectra in
these two cases \cite{morf,rh,rr,md}.

I thank R. Morf for helpful correspondence. My research is
supported by the NSF, under grant no.\ DMR-98-18259.


\vspace*{-5mm}


\begin{references}


\vspace*{-15mm}


\bibitem{willett} R.L. Willett, {\it et al.}, \prl {\bf 59}, 1779
(1987).

\bibitem{spj} V.W. Scarola, K. Park, and J.K. Jain, Nature
{\bf 406}, 863 (2000), and on the web: cond-mat/0012030.

\bibitem{jain}J.K.~Jain, \prl {\bf 63}, 199 (1989); \prb {\bf
40}, 8079 (1989); {\it ibid.} {\bf 41}, 7653 (1990).

\bibitem{hlr}B.I.~Halperin, P.A.~Lee, and N.~Read, \prb {\bf 47}, 7312
   (1993).

\bibitem{bcs}
J. Bardeen, L.N. Cooper, and J.R. Schrieffer, Phys. Rev. {\bf
106}, 162 (1957); {\bf 108}, 1175 (1957).

\bibitem{mr}
G. Moore and N. Read, Nucl.\ Phys.\ B{\bf 360}, 362 (1991).

\bibitem{hr}
F.D.M.~Haldane and E.H.~Rezayi, \prl {\bf 60}, 956, 1886 (E)
(1988).

\bibitem{gww}
M.~Greiter, X.-G.~Wen and F.~Wilczek, \prl {\bf 66}, 3205 (1991);
Nucl. Phys. B{\bf 374}, 567 (1992).

\bibitem{read94}N.~Read, Semicond. Sci. Technol. {\bf 9}, 1859 (1994)
   [=cond-mat/9501090].

\bibitem{g}S.M.~Girvin, in {\it The Quantum Hall Effect},
edited by R.E.~Prange and S.M.~Girvin (Second Edition,
Springer-Verlag, New York, 1990).

\bibitem{gm}S.M.~Girvin and A.H.~MacDonald, \prl {\bf 58}, 1252 (1987).

\bibitem{read87}N.~Read, Bull. Am. Phys. Soc, {\bf 32}, 923 (1987);
\prl {\bf 62}, 86 (1989).

\bibitem{zhk}S.C.~Zhang, T.H.~Hansson, and S.~Kivelson, \prl {\bf 62}, 82
    (1989).

\bibitem{eisenstein} J.P. Eisenstein {\it et al.}, \prl {\bf 61},
997 (1988).

\bibitem{belkhir}
L. Belkhir, X.-G. Wu, and J.K. Jain, \prb {\bf 48}, 15245 (1993).

\bibitem{morf}
R. Morf, \prl {\bf 80}, 1505 (1998).

\bibitem{rh}
E.H. Rezayi and F.D.M. Haldane, \prl {\bf 84}, 4685 (2000).

\bibitem{jim} J.P. Eisenstein {\it et al.}, Surface Sci. {\bf 229}, 31
(1990).

\bibitem{pan} W. Pan {\it et al.}, \prl {\bf 83}, 3530 (1999).

\bibitem{rr} E. Rezayi and N. Read, \prl {\bf 72}, 900 (1994);
{\it ibid.}, {\bf 73}, 1052 (C) (1994).

\bibitem{jainprlcom} J.K. Jain, \prl {\bf 73}, 1051 (C) (1994).

\bibitem{md} R. Morf and N. d'Ambrumenil, \prl {\bf 74}, 5116
(1995).

\bibitem{cooper} L.N. Cooper, Phys. Rev. {\bf 104}, 1189 (1956).

\bibitem{fetter} A.L. Fetter and J.D. Walecka, {\it Quantum Theory
of Many-Particle Systems}, (McGraw-Hill, New York, 1971), p.\ 23.

\bibitem{ll} L.D. Landau and E.M. Lifshitz, {\it Quantum
Mechanics (Nonrelativistic Theory)}, (Pergamon Press, Oxford,
1977), p. 435.

\bibitem{sm}R.~Shankar and G.~Murthy, \prl {\bf 79}, 4437 (1997);
    cond-mat/9802244.

\bibitem{dhlee}D.-H.~Lee, \prl {\bf 80}, 4745 (1998).

\bibitem{ph}V.~Pasquier and F.D.M.~Haldane, Nucl. Phys. B {\bf 516}, 719 (1998).

\bibitem{read98}N. Read, \prb {\bf 58}, 16262 (1998).

\end{references}
\end{document}